\documentclass[aps]{revtex4}

\newcommand{\leftsuperindex}[2]{  { {}^{\mbox{\tiny{$(#2)$}}} \! #1 } }
\newcommand{\three}[1]{ \leftsuperindex{#1}{3} }
\newcommand{\four}[1]{ \leftsuperindex{#1}{4} }
\newcommand{\func}{f}
\newcommand{\constantfield}{\gamma}
\newcommand{\Z}{Z}
\newcommand{\Zer}{\chi}

\usepackage{amsmath, amssymb, amsfonts}

\begin{document}


\title{A generalization of the Zerilli master variable for a dynamical spherical spacetime}

\author{David Brizuela}
\affiliation{Fisika Teorikoa eta Zientziaren Historia Saila, UPV/EHU, 644 P.K., 48080 Bilbao, Spain}
\email{david.brizuela@ehu.eus}


\begin{abstract}
The evolution of polar perturbations on a spherical background
spacetime is analyzed. The matter content is assumed to be a massless scalar field.
This provides a nontrivial dynamics to the background and the linearized
equations of motion become much more involved than in the vacuum case.
The analysis is performed in a Hamiltonian framework, which makes explicit the dynamical role of each of the variables.
After performing a number of canonical transformations,
it is possible to completely decouple the different perturbative degrees of freedom into
constrained, pure-gauge and gauge-invariant variables. In particular, two master variables
are obtained: one corresponding to the polar mode of the gravitational wave,
whereas the other encodes the complete physical information about the perturbative
matter degree of freedom. The evolution equations for these master variables
are obtained and simplified.
\end{abstract}

\maketitle

\section{Introduction}

The study of linearized perturbations of known background solutions
of general relativity has had very important contributions
to our current understanding of different gravitational scenario.
As in other field theories with constraints, one of the main problems
of this approach is the identification of physical degrees of freedom.
There are usually two approaches one can follow for such a purpose.
On the one hand, it is possible to impose convenient gauge-fixing conditions
and work on a certain gauge. On the other hand, one can construct
gauge-invariant variables so that any physical result is unambiguous
and valid in any gauge.

The Hamiltonian formalism of general relativity gives a very clear
and transparent notion of the gauge dependence. In particular, the Hamiltonian
is a linear combination of first-class constraints, which are the generators of
gauge transformations. In the context of perturbation theory, such a Hamiltonian formalism was pioneered by Moncrief 
\cite{Monc74} to study the nonspherical perturbations of
the Schwarzschild black hole. In that reference it was shown that, for this solution,
it is possible to perform several canonical transformations explicitly in such a way that
the initial twelve perturbative variables (the six components of the perturbed spatial
metric in combination with their corresponding conjugate momenta)
are reorganized into two physical pairs, which encode the complete
physical information of the gravitational wave, and four gauge pairs.
In each gauge pair, one of the variables is constrained to vanish on shell,
whereas its conjugate variable is nonphysical. Furthermore, the two physical pairs obey unconstrained
evolution equations and are equivalent to the Regge-Wheeler \cite{ReWh57} and Zerilli \cite{Zeri70}
master variables. In this way, the physical degrees of freedom are explicitly
decoupled from the gauge degrees of freedom and the dynamical behavior of
the system is completely described by the two physical pairs.
This technique was also applied to other specific solutions of Einstein equations
like Reissner-Nordstr\"om \cite{Monc74b,Monc75}, Oppenheimer-Snyder \cite{CPM78}
or Friedmann-Robertson-Walker \cite{Gund93, Lan94}.

Regarding spherically symmetric background metrics, it is well known
that the perturbations can be classified into two different sectors
(axial and polar) depending their polarity. At linear order these
two sectors decouple, but at second and higher orders they
interact as, for instance, the coupling between two axial modes
give rise to both axial and polar modes \cite{PBMT10}. For the Schwarzschild metric
the Regge-Wheeler \cite{ReWh57} and Zerilli \cite{Zeri70} variables
encode respectively the axial and polar physical degrees of freedom.
Both are master scalars since they obey unconstrained evolution equations
and the complete perturbed metric can be reconstructed in terms of them.
These two variables were initially defined on a fixed perturbative
gauge but, as commented above, were later obtained by Moncrief on a
generic gauge.

A very convenient framework to study perturbations around generic (possibly dynamical)
spherically symmetric spacetimes was presented by Gerlach and Sengupta \cite{GeSe79, GuMa00}.
This is a very geometrical framework, where the four dimensional manifold
is decomposed as the product between a two dimensional Lorentzian manifold
 with boundary and the unit two-sphere.
The construction of perturbative gauge-invariant variables is explicitly performed
and, for the axial case, a master scalar variable is constructed.
This master scalar obeys an unconstrained wave equation and can be coupled
to any kind of matter. Therefore, it can be considered as the generalization
of the Regge-Wheeler variable to dynamical spacetimes.
On the contrary, for the polar sector, there is no known master variable
valid for any spherically symmetric background. Nonetheless, some particular
results for specific metrics have been obtained. For instance,
on a vacuum background, the gauge-invariant combinations of the linearized
stress-energy tensor have also been included in \cite{MaPo05,NaRe05}.
On the other hand, in Ref. \cite{SaTi00} a polar master scalar was
defined, for a vacuum background, which was later generalized
to nonlinear electrodynamics \cite{MoSa03} for any background solution.

The present paper is motivated by the search of a unique gauge-invariant
master variable encoding the polar gravitational wave for any matter model.
In principle it is not clear that it can be constructed but,
apart from its conceptual relevance, such a polar master variable would
be of great use for several applications. For instance the numerical resolution of
an unconstrained equation is, in principle, much easier and precise
than the resolution of a coupled system of several equations which, 
in addition, are subjected to certain constraints. In addition, 
it could be particularly useful in the very hard problem of
matching of polar gravitational waves through moving surfaces, for
instance the surface of a supernova explosion \cite{Sei90, MaGu01, MMR07}. This problem becomes
even harder beyond first-order perturbation theory \cite{Muk00, Mar05, BMSK10, ReVe14}.

In a previous paper \cite{BrMa09} the Gerlach-Sengupta axial master
scalar was reobtained , making use of the Hamiltonian techniques explained above,
for a dynamical spherical background spacetime with a matter content of a scalar field.
Here the analysis of that paper is reproduced for the polar sector, as an example
of a dynamical spacetime for which no polar master variable is known.

The rest of the paper is organized as follows. In Sec. \ref{sec:hamiltonian} the
Hamiltonian framework for linearized perturbations on a generic background
is briefly reviewed. Section \ref{sec:spherical} introduces the notation and the
equations of motion corresponding to the particular background we will be dealing
with: a spherically symmetric spacetime with a massless scalar field.
In Sec. \ref{sec:polar} the polar part of the perturbative variables are
decomposed into tensor spherical harmonic and a number of canonical
transformations are performed in order to decouple the gauge and the physical
degrees of freedom. In Sec. \ref{sec:evolution} the evolution equation
for the master variables are presented and simplified. Finally, Sec. \ref{sec:conclusions}
discusses the main conclusions.


\section{General relativistic perturbation theory on a Hamiltonian framework}\label{sec:hamiltonian}

\subsection{Hamiltonian framework for general relativity}

Let us assume general relativity coupled with a massless scalar field $\Phi$.
In order to perform a Hamiltonian analysis of this system, the usual
$3+1$ decomposition of the spacetime is performed. Greek indices will be used
for four-dimensional objects and Latin indices for three-dimensional ones.
It is possible to choose coordinates $(t,x^i)$
adapted to the foliation so that the three-dimensional metric $g_{ij}$, defined by projecting
the four-dimensional one $\four{g}_{\mu\nu}$ to the spatial slices, is given as
\begin{equation}
g_{ij}:=\four{g}_{ij}.
\end{equation}
Furthermore, the lapse function $\alpha$  and the shift vector $\beta_i$
are defined as follows,
\begin{equation} \label{lapse&shift}
\alpha^{-2}:= - \four{g}^{tt} , \qquad
\beta_i := \four{g}_{ti}. \qquad
\end{equation}

The action for the system under consideration is then given by,
\begin{equation} \label{action}
{\cal S} = \int dt \int d^3x \left( 
\Pi^{ij} g_{ij,t} 
+ \Pi \Phi_{,t}- \alpha {\cal H} - \beta^i {\cal H}_i \right),
\end{equation}
where $\Pi^{ij}$, which is related to the extrinsic curvature,
is the conjugate momentum of the spatial metric $g_{ij}$ whereas
$\Pi$ is the conjugate momentum of the scalar field.
As it is well known, the lapse and the shift are Lagrange multipliers
associated to the Hamiltonian ${\cal H}$ and momentum constraint
${\cal H}_i$ respectively. These take the following form in
terms of the basic variables:
\begin{eqnarray}
{\cal H} &=& \frac{1}{\mu_g}
\left[\Pi^{ij}\Pi_{ij}-\frac{1}{2}\left(\Pi^l{}_l\right)^2\right]
     - \mu_g \three{R} 
     + \frac{1}{2} \left( \frac{\Pi^2}{\mu_g}
                  +  \mu_g g^{ij} \Phi_{,i}\Phi_{,j}
\right), \\
{\cal H}_i &=&  -2 D_j\Pi_i{}^j + \Pi \Phi_{,i},
\end{eqnarray}
where $\mu_g:=\sqrt{\det g_{ij}}$ and
$D_j$ is the covariant derivative associated to $g_{ij}$.

\subsection{Linearized perturbations}

Let us begin by defining a one-parameter family of spacetimes
$({\cal M}(\varepsilon), \tilde g_{\mu\nu}(\varepsilon))$, where $\varepsilon$
is a dimensionless parameter. The $\varepsilon=0$ member of
the family is referred as the background spacetime.
The idea behind linear perturbation theory is to perform
a linearization around a background, which is a known exact
solution of the Einstein equations. This is done by performing a Taylor
expansion on the parameter $\varepsilon$ of all quantities that appear in the equations.
Then, terms higher than second order in $\varepsilon$ are dropped.
For convenience, we define the operator
\begin{equation}
\delta^n F(\varepsilon):=\left.\frac{d^n F(\varepsilon)}{d\varepsilon^n}\right|_{\varepsilon=0},
\end{equation}
and, making use of it, introduce the following notation for the perturbative variables:
\begin{eqnarray}\nonumber
C := \delta \alpha , \quad &\qquad&
B^i := \delta(\beta^i) , \\
h_{ij} := \delta(g_{ij}) , &\qquad& 
p^{ij} := \delta(\Pi^{ij}) .
\label{perturbative_variables}
\end{eqnarray}

As it was shown in Ref. \cite{Taub, Monc74}, the second variation of
the action,
\begin{equation}\label{effaction}
\frac{1}{2}\delta^2{\cal S} =
\int dx^4 \left[ 
p^{ij} h_{ij,t} 
+ \delta\Pi\,\, \delta\Phi_{,t}
- C \delta({\cal H})
- B^i \delta({\cal H}_i)
-\frac{\alpha}{2} \delta^2({\cal H})
- \frac{\beta^i}{2} \delta^2({\cal H}_i)
\right] ,
\end{equation}
provides an action functional for the linearized perturbations.
(The explicit form of the first and second variations of
the Hamiltonian and momentum constraints can be found in \cite{BrMa09}.)
That is, the variation of this last action with respect to
different perturbative variables (\ref{perturbative_variables})
gives the linearized Einstein equations. In particular,
the variation with respect to the perturbation of
the lapse $C$ and the shift $B^i$ leads to
the constraints obeyed by these linearized variables,
\begin{equation}\label{constrainteq}
\delta({\cal H})=0, \qquad \delta({\cal H}_i)=0.
\end{equation}
These are first-class constraints and, therefore, the generators
of the perturbative gauge transformations. The idea of this
paper is to perform canonical transformations of the perturbative
variables so that each of these four constraints is simply expressed
as one of the new variables. In this way, the different dynamical sectors would get decoupled.
From the eighteen variables under consideration [the sixteen that appear
in (\ref{perturbative_variables}), in combination with the two functions
$(\delta\Phi,\delta\Pi)$ encoding the scalar degree of freedom], four of them
would be constrained to vanish. The four canonical conjugate variables of
the constrained ones will be pure gauge, whereas the four functions $(C,B^i)$
are Lagrange multipliers with vanishing conjugate momentum
and thus non-dynamical. Finally, the remaining six variables will be automatically
gauge invariant. These latter six variables stand for the three physical
degrees of freedom of this problem: two corresponding to the gravitational wave
and one to the perturbations of the matter scalar field.
This separation between the gauge and the physical sectors is not obvious and
can only be performed for highly symmetric backgrounds.

\section{Spherical background}\label{sec:spherical}

We will consider a spherically symmetric background, which can be decomposed
as ${\cal M}=M^2\times S^2$, where $M^2$ is a two dimensional background
with boundary and $S^2$ the unit two-sphere. Arbitrary coordinates
$(t,\rho)$ on $M^2$ and spherical coordinates $x^a=(\theta,\phi)$
on $S^2$ are chosen. The lower-case Latin indices stand for coordinates on the two-sphere.
Due to the symmetry, the lapse can only be a function of the coordinates
on $M^2$, that is, $\alpha=\alpha(t,\rho)$; whereas the shift vector has vanishing
angular components $\beta^i=(\beta(t,\rho),0,0)$. In this way, the four-dimensional
background metric takes the following form:
\begin{equation}
(ds^2)_4 = -\alpha^2 dt^2 + a^2(d\rho +\beta dt)^2 + r^2 d\Omega^2.
\end{equation}

Furthermore the following three variables are defined, which completely encode
the information contained in the background moments $\Pi^{ij}$ and $\Pi$:
\begin{equation}
\Pi_1 :=
\frac{a^2 \Pi^{\rho\rho}}{\mu_g} , \qquad 
\Pi_2 :=
\frac{2r^2 \Pi^{\theta\theta}}{\mu_g} ,\qquad
\Pi_3 :=
\frac{\Pi}{\mu_g}.
\end{equation}

For completeness, here the symmetry-reduced background constraints are provided,
\begin{eqnarray}\label{sphericalconstraint1}
\frac{\cal H}{\mu_g} &=&
\Pi_1\left(\frac{\Pi_1}{2}-\Pi_2\right)
- \three{R} +\frac{1}{2} \left({\Pi_3}^2+{\Phi'}^2\right)
= 0 ,
\\\label{sphericalconstraint2}
\frac{1}{a}\frac{{\cal H}_\rho}{\mu_g} &=&
-\frac{2}{r^2}(r^2\Pi_1)'+\frac{2r'}{r}\Pi_2
+ \Pi_3 \Phi' = 0 ,
\end{eqnarray}
where prime stands for the derivative with respect to $\rho$ divided by
the function $a$:
\begin{equation}
f'=\frac{f_{,\rho}}{a}.
\end{equation}
These constraints generate the following evolution equations:
\begin{eqnarray}
\frac{1}{\alpha}\left[a_{,t}-(\beta a)_{,\rho}\right] &=& \frac{a}{2} \left(\Pi_1-\Pi_2\right) , \\\label{Pi1}
\frac{1}{\alpha}(r_{,t}-\beta r_{,\rho}) &=& -\frac{r}{2} \Pi_1 , \\
\frac{1}{\alpha}(\Phi_{,t}-\beta \Phi_{,\rho}) &=& \Pi_3\\\label{pi1dot}
\frac{1}{\alpha}\left( \Pi_1{}_{,t} -\beta \Pi_1{}_{,\rho } \right)
&=& 
\frac{3\Pi_1^2}{4} 
+ \frac{1}{r^2}
- \frac{r'}{r}\frac{(\alpha^2 r)'}{\alpha^2 r}
+ \frac{1}{4}\left(\Pi_3^2+{\Phi'}^2\right), \\\label{pi2dot}
\frac{1}{\alpha}\left( \Pi_2{}_{,t} - \beta \Pi_2{}_{,\rho} \right)
&=&
\frac{1}{2}(\Pi_1^2+\Pi_2^2-\Pi_1\Pi_2)
+ \frac{2\alpha'r'}{\alpha r}
- \frac{2(\alpha r)''}{\alpha r}
+ \frac{1}{2}\left(\Pi_3^2-{\Phi'}^2\right), \\
\frac{1}{\alpha}(\Pi_3{}_{,t}-\beta \Pi_3{}_{,\rho}) &=& 
\frac{\Pi_3(\Pi_1+\Pi_2)}{2}
+\frac{(\alpha r^2\Phi')'}{\alpha r^2}.
\end{eqnarray}
The case studied previously by Moncrief is vacuum in Schwarzschild coordinates, which
is recovered by choosing $\Phi=\Pi_1=\Pi_2=\Pi_3=0$. This restriction greatly simplifies
the above equations of motion and in particular both the Hamiltonian (\ref{sphericalconstraint1})
and momentum constraints (\ref{sphericalconstraint2}) are trivially fulfilled.

\section{Polar perturbations}\label{sec:polar}
\subsection{Expansion in harmonics}

In order to take advantage of the background spherical symmetry we will make
use of the tensor spherical harmonics in order to decompose the perturbations.
Properties and precise definitions of the harmonics that will be used here
can be found in Ref. \cite{BMM06}. Following their behavior under a parity
transformation, different tensor spherical harmonics can be divided into
two groups: axial harmonics, with a polarity $(-1)^{l+1}$, and polar
harmonics, with a polarity $(-1)^l$. Since at a linearized level these two
groups of harmonics decouple, it is possible to consider the axial and
the polar problem independently. As commented in the introduction, in Ref. \cite{BrMa09} the axial case
was developed, whereas here we will focus on the polar case. The decomposition
of the polar part of the different perturbations is given as follows,
\begin{eqnarray}\label{dec1}
h_{ij}dx^idx^j &=&\sum_{l=0}^\infty \sum_{m=-l}^l a^2 (H_2)_l^m Y_l^m d\rho^2
+ 2 (h_1)_l^m d\rho \, Z_l^m{}_{a}dx^a
+ r^2\left[ K_l^m \gamma_{ab} Y_l^m
+ G_l^m Z_l^m{}_{ab} \right] dx^a dx^b ,\\
\frac{1}{\mu_g}p_{ij}dx^idx^j &=&\sum_{l=0}^\infty \sum_{m=-l}^l
a^2 (P_H)_l^m Y_l^m d\rho^2 + 2 (P_h)_l^m d\rho \, Z_l^m{}_{a}dx^a
+ r^2 \left[ (P_K)_l^m \gamma_{ab} Y_l^m + (P_G)_l^m Z_l^m{}_{ab} \right] dx^a dx^b,\\
C &=& \sum_{l=0}^\infty \sum_{m=-l}^l\frac{-\alpha}{2} (H_0)_l^m Y_l^m,\\
B_i dx^i &=& \sum_{l=0}^\infty \sum_{m=-l}^l (H_1)_l^m \, Y_l^m d\rho + (h_0)_l^m \, Z_l^m{}_{a}dx^a ,\\
\frac{1}{\mu_g}\delta\Pi &=& \sum_{l=0}^\infty \sum_{m=-l}^l \hat p_l^m Y_l^m , 
\\\label{dec6}
\delta\Phi &=& \sum_{l=0}^\infty \sum_{m=-l}^l \varphi_l^m Y_l^m ,
\end{eqnarray}
where the notations by Regge-Wheeler and Moncrief have been followed
for the different harmonic coefficients.

Since all perturbations are decoupled at linear order, from here on,
the $(l,m)$ labels from the harmonic coefficients and the harmonic tensors
will be removed.
In addition, we define the following shortening for the sum that
appears in all decompositions above,
\begin{equation}
\sum_{l,m}:=\sum_{l=0}^\infty \sum_{m=-l}^l.
\end{equation}

\subsection{Effective action}

In order to obtain the effective action in terms of the harmonic coefficients,
decomposition (\ref{dec1}-\ref{dec6}) is introduced in expression
(\ref{effaction}). The angular part can be integrated by
making use of the properties of the tensor spherical harmonics as
shown in appendix of Ref. \cite{BrMa09}. In this way, it is easy to obtain
the following form for the effective action for the polar part of the linearized perturbations:
\begin{eqnarray}\label{effectiveaction}
\frac{1}{2}\left(\delta^2{\cal S}\right)^{\rm polar} &=&
\int dx^4 \left[ 
p^{ij} h_{ij,t} 
+ \delta\Pi\,\, \delta\Phi_{,t}
- C \delta({\cal H})
- B^i \delta({\cal H}_i)
-\frac{\alpha}{2} \delta^2({\cal H})
- \frac{\beta^i}{2} \delta^2({\cal H}_i)
\right]^{\rm polar} \\\nonumber
&=&\sum_{l,m}\int dt
\int d\rho \left[\, p_1 h_{1,t} + p_2 H_{2,t} + p_3 K_{,t} + p_4 G_{,t} + p \varphi_{,t} \right]
+\int dt\, \{ F_0[-\alpha H_0/2]
+ F_1[H_1] + F_2[h_0]\} + ...,
\end{eqnarray}
where the dots stand for terms coming from the second perturbation of
the background constraints,
which do not enter the gauge transformations,
and the functionals  $(F_0,F_1,F_2)$ will be defined below.
The conjugate momenta are related to the harmonic coefficients given by
the expansions (\ref{dec1}-\ref{dec6}) in the following way,
\begin{eqnarray}
p_1 &=& \frac{2l(l+1)}{a} P_h^*, \\
p_2 &=& ar^2 P_H^* , \\
p_3 &=& 2ar^2 P_K^* , \\
p_4 &=& \lambda a r^2 P_G^* , \\
p &=& a r^2 \hat{p}^* ,
\end{eqnarray}
where the star stands for complex conjugate and $$\lambda:=\frac{1}{2}\frac{(l+2)!}{(l-2)!},$$ has been defined.

The polar part of the harmonic decomposition of the
linearized constraints is given by,
\begin{eqnarray*}
\delta[{\cal H}] &=& \sum_{l,m}\mu_g Y \left\{ 
H_2 \left[ \Pi_1(\Pi_1-\Pi_2)-\frac{l^2+l+2}{r^2}-\frac{\cal H}{2\mu_g}\right]
-2 H_2{}'\frac{r'}{r}
+ \frac{p_2}{ar^2} (\Pi_1-\Pi_2)
+ \frac{1}{2} K \left[ -\Pi_1^2-\Pi_3^2+\Phi'{}^2\right]
 \right. \nonumber \\ 
&-& \frac{p_3}{ar^2}\Pi_1
-\left[\three{R}+\frac{(l-1)(l+2)}{r^2}\right]K
\left.
+\frac{2}{r^3}(r^3K{}')' 
+\Phi'\varphi'+\frac{p}{ar^2}\Pi_3
+\frac{2l(l+1)}{r^3}(ra^{-1}h_1)'
-\frac{\lambda}{r^2} G \right\}
, \\
\frac{1}{a}\delta[{\cal H_\rho}]&=&\sum_{l,m} \mu_g Y \left\{ 
\!-\frac{2(a^{-1}p_2)'}{r^2}\!+\!\frac{p_1}{r^2}\!+\!\frac{2p_3}{ar^2}\frac{r'}{r}\!+\!\frac{p}{ar^2}\Phi'
\!-\!H_2{}'\Pi_1
\!-\!\frac{2}{r^2}(r^2\Pi_1)'H_2
\!+\!\frac{l(l+1)}{ar^2}\Pi_2h_1
\!+\!\frac{\Pi_2}{r^2}(r^2K)' \!+\!\Pi_3\varphi'
\!\right\}\!, \\
\delta[{\cal H}_a]^{\rm polar} \!\!\! &=& \!\!\!\sum_{l,m} \mu_g Z_{a} \left\{ 
\frac{-1}{l(l+1)}\frac{(r^2p_1)'}{r^2}-\frac{p_3}{ar^2}+\frac{2}{l(l+1)}\frac{p_4}{ar^2}
+\frac{(l-1)(l+2)}{2} \Pi_2 G 
+\Pi_1 H_2-\frac{2}{r^2}\left(\frac{r^2\Pi_1 h_1}{a}\right)'
+\Pi_3\varphi
\right\}.
\end{eqnarray*}
With these relations at hand, the three generators
of polar gauge transformations can be written as
\begin{eqnarray}
F_0[\func ]&=&\int dx^3 \func \, Y \delta[{\cal H}] , \\
F_1[\func ]&=&\int dx^3 \func  \, Y \frac{1}{a}\delta[{\cal H}_\rho] , \\
F_2[\func ]&=&\int dx^3 \func  \, Z_{a} \frac{1}{r^2}\gamma^{ab} \delta[{\cal H}_b]^{\rm polar},
\end{eqnarray}
which act on any smooth arbitrary scalar field $\func $. It is possible to
calculate the Poisson brackets between different generators,
\begin{eqnarray}
\left\{F_0[\func_1{} ],F_0[\func_2]\right\} &=& \int d\rho \, ar^2(\func_1\func_2'-\func_1'\func_2)\frac{1}{a}\frac{{\cal H}_\rho}{\mu_g} , \\
\left\{F_0[\func_1],F_1[\func_2]\right\} &=& \int d\rho \, a \func_1 \left(r^2 \func_2 \frac{\cal H}{\mu_g}\right)' , \\
\left\{F_0[\func_1],F_2[\func_2]\right\} &=& - l(l+1) \int d\rho \, a \func_1\func_2 \frac{\cal H}{\mu_g} , \\
\left\{F_1[\func_1],F_1[\func_2]\right\} &=& \int d\rho \, ar^2 (\func_1\func_2'-\func_1'\func_2)\frac{1}{a}\frac{{\cal H}_\rho}{\mu_g} , \\
\left\{F_1[\func_1],F_2[\func_2]\right\} &=& 0 , \\
\left\{F_2[\func_1],F_2[\func_2]\right\} &=& l(l+1) \int d\rho \, a (\func_1\func_2'-\func_1'\func_2)\frac{1}{a}\frac{{\cal H}_\rho}{\mu_g},
\end{eqnarray}
all vanishing on-shell, which confirms that they are first-class constraints.

\subsection{Canonical transformations: gauge-invariant variables}

Moncrief isolated the Zerilli variable after two canonical transformations
on the four pairs $(h_1, p_1),\,(H_2,p_2),\,(K,p_3),\,(G,p_4)$. Here
the additional pair $(\varphi,p)$ for the scalar field is also present;
and the fact that the background is dynamical makes the problem harder.
We will instead proceed in five steps, to clarify the role of each step
and simplify the computations. In particular
we will first eliminate the two gauge degrees of freedom related to the
momentum constraint (which are rather trivial and very similar to
the axial case) and then remove the gauge degree associated with
the Hamiltonian constraint, the nontrivial step of this computation.

There are many possible transformations that implement this program,
but we would like them to obey certain minimal criteria.
First, they should be algebraic transformations so that they
do not involve any integration in the process and can be performed
explicitly. Second, they should not require dividing by any
background object that could vanish, in particular one of the
background momenta $(\Pi_1,\Pi_2,\Pi_3)$. And third, in order to obtain
a generalization of the Zerilli master variable, all transformations
should be well defined in the vacuum limit, that is, when taking
vanishing values for the scalar field and its perturbations. The full transformation
that will be proposed here completely fulfills the first and third criterion, but
it is unclear whether it also satisfies the second one, as will explained below.

The first canonical transformation is motivated by the Gerlach and
Sengupta choice of gauge invariants \cite{GeSe79},
\begin{eqnarray}\label{trans1first}
k_1 &=& K + \frac{l(l+1)}{2}G 
- \frac{2r'}{r}\left(a^{-1}h_1-\frac{r^2}{2}G'\right) , \\
k_2 &=& H_2 - 2\left(a^{-1}h_1-\frac{r^2}{2}G'\right)' , \\
k_3 &=& G , \\
k_4 &=& a^{-1} h_1 - \frac{r^2}{2}G' , \\
k_5 &=& \varphi - \left(a^{-1}h_1-\frac{r^2}{2}G'\right) \Phi' ,
\end{eqnarray}
which requires the canonical momenta
\begin{eqnarray}
\pi_1 &=& p_3 , \\
\pi_2 &=& p_2 , \\
\pi_3 &=& p_4 -\frac{l(l+1)}{2}p_3 -\frac{1}{2}a(r^2p_1)' , \\
\pi_4 &=& a p_1 - 2a (a^{-1}p_2)'+\frac{2r'}{r}p_3+p\,\Phi' , \\\label{trans1last}
\pi_5 &=& p .
\end{eqnarray}
In terms of these new variables, the components of the perturbed
momentum constraints are written as
\begin{eqnarray}
\frac{1}{a}\delta[{\cal H}_\rho] &=&\sum_{l,m} \mu_g Y \left\{
\frac{\pi_4}{ar^2}
+\Pi_2 \frac{(r^2k_1)'}{r^2}
-\Pi_1 k_2'
-2k_2 \frac{(r^2\Pi_1)'}{r^2}
-\frac{2}{r^2}(r^2\Pi_1 k_4')'
+k_4' \frac{1}{a}\frac{{\cal H}_\rho}{\mu_g}
\right. \nonumber \\ &+&\left.
\frac{l(l+1)}{r^2}\Pi_2 \left(k_4-k_3 rr'\right)
+k_4 \Pi_3 \Phi''
+k_4\frac{\Pi_2}{r^2}(r^2)''
+\Pi_3 k_5' \right\}
\label{deltaHsubrho}
, \\
\delta[{\cal H}_a]^{\rm polar} &=&\sum_{l,m} \mu_g Z_{a} \left\{
\frac{2\pi_3}{l(l+1)ar^2}
+k_2\Pi_1 
+\Pi_3 k_5
-\frac{(r^2)'}{r^2}\Pi_2k_4
+ \frac{\lambda}{l(l+1)} \Pi_2 k_3
-\frac{1}{r^2}(\Pi_1 r^4 k_3')' 
+ \frac{k_4}{a}\frac{{\cal H}_\rho}{\mu_g} \right\}.
\label{deltaHpolar}
\end{eqnarray}
The explicit form of the perturbation of the
Hamiltonian constraint $\delta[{\cal H}]$ in terms of these
variables is not displayed because it is a very lengthy expression and does not
contribute in any way to the present discussion.

A second canonical transformation is performed, which converts the
momentum constraints into canonical variables. Because of the
requirements we want to impose in all our transformations,
only the momenta $\pi_4$ and $\pi_3$ can replace the constraints
(\ref{deltaHsubrho}) and (\ref{deltaHpolar}) respectively,
\begin{eqnarray}
\bar{\pi}_1 &=& \pi_1 - a r^2 (\Pi_2 k_4)' , \\
\bar{\pi}_2 &=& \pi_2 +\frac{l(l+1)}{2} a r^2 \Pi_1 k_3 + a r^2 \Pi_1 k_4'- a (r^2\Pi_1)' k_4 , \\
\bar{\pi}_3 &=&\frac{1}{2}{l(l+1)ar^2} \left(\frac{\delta[{\cal H}_a]^{\rm polar}}{\mu_g Z_{a}}\right)= \pi_3 + ... , \\
\bar{\pi}_4 &=& r^2 \left(\frac{\delta[{\cal H}_\rho]}{\mu_g Y }\right)= \pi_4 + ... , \\
\bar{\pi}_5 &=& \pi_5 + \frac{l(l+1)}{2} a r^2 \Pi_3 k_3 - a (r^2 \Pi_3 k_4)'.
\end{eqnarray}
The division by the tensor harmonics must be understood just
as removing them, as well as the summation symbol, from the above expressions
(\ref{deltaHsubrho}--\ref{deltaHpolar}).
These last transformations for the momenta do not affect the position variables,
\begin{eqnarray}
\bar{k}_1 &=& k_1 , \\
\bar{k}_2 &=& k_2 , \\
\bar{k}_3 &=& k_3 , \\
\bar{k}_4 &=& k_4 , \\
\bar{k}_5 &=& k_5 .
\end{eqnarray}
In terms of these last variables, the perturbative constraints take the following simpler form,
\begin{eqnarray}
\delta[{\cal H}] &=& \sum_{l,m}\mu_g Y \left\{ 
-\Pi_1 \frac{\bar{\pi}_1}{ar^2} 
+(\Pi_1-\Pi_2)\frac{\bar{\pi}_2}{ar^2}
+\Pi_3\frac{\bar{\pi}_5}{ar^2}
+\frac{2}{r^3}(r^3\bar{k}_1')'
-\frac{(r^2)'}{r^2}\bar{k}_2'
+\bar{k}_5'\Phi'
\right. \nonumber\\
&+&\left.
\left[\Pi_1(\Pi_2-\Pi_1)-\frac{(l-1)(l+2)}{r^2}-\Pi_3^2\right]\bar{k}_1
\label{intermediatedeltaH}
-\left[\Pi_1(\Pi_2-\Pi_1)
+\frac{(l-1)(l+2)+4}{r^2}\right]\bar{k}_2
\right\}
\label{finalHperturbed} , \\
\frac{1}{a}\delta[{\cal H}_\rho] &=&\sum_{l,m} \mu_g Y \frac{\bar{\pi}_4}{ar^2} , \\
\delta[{\cal H}_a]^{\rm polar} &=& \sum_{l,m}\mu_g Z_{a} \frac{2}{l(l+1)}\frac{\bar{\pi}_3}{ar^2} .
\end{eqnarray}
The gauge freedom contained in the perturbed momentum constraints has been
fully isolated. The variables $\bar{k}_3$ and $\bar{k}_4$ are
gauge dependent and non-dynamical because their conjugate momenta
$\bar{\pi}_3$ and $\bar{\pi}_4$ are constrained to vanish.
We are left with a system of three degrees of freedom $(\bar{k}_1,
\bar{k}_2, \bar{k}_5)$ and the single constraint (\ref{intermediatedeltaH}).

Following the same procedure, at this point one should make another
canonical transformation and convert the Hamiltonian constraint into
one of the variables. Because of the first criterion we want to impose,
we can not convert any of the variables $\{\bar{k}_1,\bar{k}_2,\bar{k}_5\}$
which appear in (\ref{intermediatedeltaH}) into the full constraint. But,
because of the second requirement, we can neither do it for any of the
momenta $\{\bar{\pi}_1,\bar{\pi}_2,\bar{\pi}_5\}$.
Therefore, the idea is to first make a transformation that removes
second-order derivatives of $\bar{k}_1$ from the constraint (\ref{intermediatedeltaH}),
so that all first derivatives of the perturbed objects can be absorbed
in a single term. We use an arbitrary constant $\constantfield$ to parameterize
the transformation,
\begin{eqnarray}\label{trans4first}
\tilde{k}_1 &=& \bar{k}_1 , \\
\tilde{k}_2 &=& \bar{k}_2 - \frac{r\bar{k}_1'}{r'}-\constantfield \bar{k}_1,\\
\tilde{k}_3 &=& \bar{k}_3 , \\
\tilde{k}_4 &=& \bar{k}_4 , \\
\tilde{k}_5 &=& \bar{k}_5 ,
\end{eqnarray}
which will introduce a first derivative of $\bar{\pi}_2$
in the Hamiltonian constraint through the transformations
of the momenta,
\begin{eqnarray}
\tilde{\pi}_1 &=& \bar{\pi}_1 +(\constantfield-1)\bar{\pi}_2
- ar\left(\frac{a^{-1}\bar{\pi}_2}{r'}\right)', \\
\tilde{\pi}_2 &=& \bar{\pi}_2 ,\\
\tilde{\pi}_3 &=& \bar{\pi}_3 , \\
\tilde{\pi}_4 &=& \bar{\pi}_4 , \\
\tilde{\pi}_5 &=& \bar{\pi}_5 .\label{trans4last}
\end{eqnarray}

In this way, the Hamiltonian constraint can be written as
a sum of a full derivative term (which will be later promoted to the polar gauge-invariant
geometric master variable) and a linear combination of
variables $\tilde{k}_i$ and $\tilde{\pi}_i$ with no derivatives,
\begin{eqnarray}
\delta[{\cal H}] &=&\sum_{l,m} \mu_g Y
\left\{\left[-\Pi_1\frac{\tilde{\pi}_2}{ar^2r'} - 2 \frac{r'}{r^2} \tilde{k}_2
+\frac{1}{r}\Phi' \tilde{k}_5 -\frac{1}{r'} V_\constantfield \tilde{k}_1\right]' - \Pi_1 \frac{\tilde{\pi_1}}{ar^2}
+ \Pi_3 \frac{\tilde{\pi_3}}{ar^2}
+ \left( \frac{r (\Pi_1)'}{r'} -\Pi_1-\Pi_2+\constantfield \Pi_1\right)\frac{\tilde{\pi}_2}{ar^2}
\nonumber\right.\\\nonumber
&+&\left.
\left[ \frac{r}{r'}  V_\constantfield'
+ V_\constantfield \frac{r^2}{2(r')^2} \left(\frac{3(r')^2}{r^2}
-\frac{1}{r^2}+\frac{\three{R}}{2}\right)
-\Pi_3^2
+(1-\constantfield)\Pi_1(\Pi_2-\Pi_1)
-\frac{l(l+1)}{r^2}(1+\constantfield)
+\frac{2}{r^2}(1-\constantfield)
\right] \tilde{k}_1
\right.\\
&-&\left.
\Phi''\tilde{k}_5 
 - \left(V+\frac{3(r')^2}{r^2}\right)\tilde{k}_2 \right\},\label{deltaHfullderivative}
\end{eqnarray}
where the background potentials,
\begin{equation}
V := \frac{1+l+l^2}{r^2} + \Pi_1(\Pi_2-\Pi_1)
         + \frac{\three{R}}{2} ,
\qquad
V_\constantfield := V + (2\constantfield-3)\frac{(r')^2}{r^2} ,
\end{equation}
have been defined.
As it has been anticipated, this clearly motivates another canonical
transformation in which the term that appears inside the full derivative
[in the first line of Eq. (\ref{deltaHfullderivative})]
replaces the canonical variable $\tilde{k}_2$. Note that using $\tilde{k}_1$ instead to replace such term would require
dividing by $V_\constantfield$, which is a background object that could vanish, whereas using  $\tilde{k}_5$ would
not provide a well defined vacuum limit.
The fourth canonical transformation takes thus the following form:
\begin{eqnarray}
\check{k}_1 &=& \tilde{k}_1 , \\\label{ztrans1}
\check{k}_2 &=& -\Pi_1\frac{\tilde{\pi}_2}{ar^2r'} - 2 \frac{r'}{r^2} \tilde{k}_2
+\frac{1}{r}\Phi' \tilde{k}_5 -\frac{1}{r'} V_\constantfield \tilde{k}_1 , \\
\check{k}_3 &=& \tilde{k}_3 , \\
\check{k}_4 &=& \tilde{k}_4 , \\
\check{k}_5 &=& \tilde{k}_5 ,\\
\check{\pi}_1 &=& \tilde{\pi}_1 - \frac{V_\constantfield}{2(r')^2} \tilde{\pi}_2, \\
\check{\pi}_2 &=& -\frac{r\tilde{\pi}_2}{2r'} ,
\\
\check{\pi}_3 &=& \tilde{\pi}_3 , \\
\check{\pi}_4 &=& \tilde{\pi}_4 , \\
\check{\pi}_5 &=& \tilde{\pi}_5 + \frac{\Phi'}{2r'}\tilde{\pi}_2 .
\end{eqnarray}
Now the Hamiltonian constraint does not contain $\check{\pi}_2$ and, in addition,
it has no explicit dependence on the constant $\gamma$. But, more
importantly, we have achieved what we were looking for: it neither contains
derivatives of $\check{k}_1$,
\begin{equation} \label{Ham4}
\delta[{\cal H}] =\sum_{l,m}\mu_g Y\left\{ D \check{k_1} 
-\Pi_1\frac{\check{\pi}_1}{ar^2}+\Pi_3 \frac{\check{\pi}_5}{ar^2}
+\check{k}_2' 
+\frac{\check{k}_2}{2}\left(\frac{V}{r'}+3r'\right)
-\check{k}_5 \left[ \Phi''+\frac{\Phi'}{2}\left(\frac{V}{r'}+3r'\right)\right]\right\}.
\end{equation}
where we have defined the background coefficient
\begin{equation}\label{background_coefficient}
\mbox{}\!D\!=\! \frac{(r^2V)'}{r r'}
+ \frac{r^2}{2(r')^2}\!\left[V\!+\!\left(\frac{r'}{r}\right)^2\right]
\!\left[\frac{l(l+1)}{r^2}+\Pi_1(\Pi_2-\Pi_1)+\three{R}\right]
\!-\! \left[2 \frac{l(l+1)}{r^2} + \Pi_3^2\right] .
\end{equation}
This fact permits us to perform the final fifth
canonical transformation, which converts the Hamiltonian
constraint into the first of the variables of the problem,
\begin{eqnarray}
Q_1 &=& D \check{k_1}-\Pi_1\frac{\check{\pi}_1}{ar^2}+\Pi_3 \frac{\check{\pi}_5}{ar^2}
+\check{k}_2' 
+\frac{\check{k}_2}{2}\left(\frac{V}{r'}+3r'\right)
-\check{k}_5 \left[ \Phi''+\frac{\Phi'}{2}\left(\frac{V}{r'}+3r'\right)\right],  \\\label{ztrans2}
Z &=& \check{k}_2 , \\
Q_3 &=& \check{k}_3 , \\
Q_4 &=& \check{k}_4 , \\
\phi &=& \check{k}_5 + \frac{\Pi_3}{ar^2D}\check{\pi}_1, \\
P_1 &=& \frac{\check{\pi}_1}{D}, \\
P_Z &=& \check{\pi}_2 + a \left(\frac{\check{\pi}_1}{aD}\right)'
-\frac{\check{\pi}_1}{2D}\left(\frac{V}{r'}+3r'\right) , \\
P_3 &=& \check{\pi}_3 , \\
P_4 &=& \check{\pi}_4 , \\
P_\phi &=& \check{\pi}_5 + \frac{\check{\pi_1}}{D}
\left[\Phi''+\frac{\Phi'}{2}\left(\frac{V}{r'}+3r'\right)\right].
\end{eqnarray}

At this point we have succeeded in separating the physical degrees of
freedom $(\Z,P_Z)$ and $(\phi,P_\phi)$ from the gauge degrees of freedom
$(Q_1,P_1)$, $(Q_3,P_3)$ and $(Q_4,P_4)$. However in this last
transformation the background object $D$ appears as denominator
and it is not clear to us whether this object can vanish or not.
In vacuum, for a Schwarzschild solution, defining
$\Lambda:= (l-1)(l+2)/2$, we have
\begin{equation}\label{deflambda}
D = \frac{1}{1-2M/r} \frac{l(l+1)}{r^3}\left(\Lambda r+3M\right),
\end{equation}
which is always positive. It is reasonable to assume that for spacetimes
close enough to Schwarzschild (though possibly dynamical),
the variable $D$ will also be positive. If this was the
case we would have succeeded in implementing to completion
the procedure while obeying the three imposed criteria.
If not, analyzing the procedure that has been followed, it seems quite
difficult to achieve the construction of gauge-invariant master variables
without dividing by a never vanishing background object for a generic background gauge.
Note that, once the canonical transformation (\ref{trans1first}--\ref{trans1last}) is performed,
there is no much freedom left in the procedure if one insists on imposing
the three criteria: performing only algebraic transformations, not dividing by a possibly vanishing
background object and having a well defined vacuum limit. More specifically, the momenta $\pi_3$ and $\pi_4$ are the only variables that 
can be used to solve for the constraints (\ref{deltaHpolar}) and (\ref{deltaHsubrho}) respectively.
This leads to the form (\ref{intermediatedeltaH}) of the linearised Hamiltonian constraint.
In that expression none of the variables can be used to solve the constraint
algebraically. Thus, next parametrized transformation (\ref{trans4first}--\ref{trans4last})
is performed in order to concentrate all derivatives in a unique full derivative.
The term inside this full derivative is then promoted to one of the basic variables.
Finally, in expression (\ref{Ham4}) the underived variable $\check k_1$ is used
to solve for the Hamiltonian constraint. Note that $\check k_5$ could also be used
to solve for that constraint, but this would not obey our third criterion about
having a well defined vacuum limit. In this case
the perturbations of the scalar field would be pure gauge
and thus the physical matter degrees of freedom would be encoded in a geometric pair.
In addition we would not get a master variable that could be consider the generalization
of the Zerilli variable.

In any case, there are other routes that one could follow to construct a polar
master gauge invariant. For instance,
one could also choose for instance a particular background
gauge with a fixed (nonzero) value of $\Pi_1$ (or $\Pi_3$)
and divide by this
moment when solving the constraint for $\check\pi_1$ (or for $\check\pi_5$).
Another alternative could be to relax the first condition about
the algebraic nature of the transformations.

\section{Evolution equations}\label{sec:evolution}

The variable $\Z$ (\ref{ztrans2})
obeys a complicated equation of motion. This section
summarizes its differential structure, and shows that it is indeed a generalization
of the Zerilli equation. In order to simplify the calculations, $(t,\rho=r)$ are
chosen as background coordinates. In addition we will take $\beta=0$ which, because of the
background evolution equation (\ref{Pi1}), implies $\Pi_1=0$.
By reversing all canonical transformations, it is straightforward to write
the variable $\Z$ in terms of the initial harmonic coefficients (\ref{dec1}--\ref{dec6}). In particular, in this
background gauge the master variable takes the following form:
\begin{eqnarray}
\Z&=&-\frac{2}{ar}H_2+\frac{1}{2a}(4 K_{,r} + 2 \Phi_{,r}\varphi-\Pi_3^2 r^2 G_{,r} )+\frac{1}{8a r}[2K+l(l+1)G]
\{12 - a^2 [4 + 4 l (1 + l) + \Pi_3^2 r^2] - (r \Phi_{,r})^2\}\nonumber\\
&+&\frac{1}{ar^2}[2 l (1 + l) + \Pi_3^2 r^2]h_1.
\end{eqnarray}
In order to get this expression, it is enough to reverse all transformations presented
in the previous section. The moments $(p_1,p_2,p_3,p)$ can be written in terms of time derivatives
of their corresponding position variable by making use of the linearised evolution equations.
Even so, note that all time derivatives disappear from the expression of the master variable
in the chosen background gauge.

Let us now present the physical evolution equations.
The equations of motion for the master gauge-invariant variables are
obtained by direct variation of the action (\ref{effectiveaction}),
\begin{eqnarray}\label{P2dot}
{P_Z}_{,t} &=& M_{11}^{(4)} P_Z^{\rm (iv)} + M_{12}^{(6)} \Z^{\rm (vi)}
+ M_{13}^{(5)} P_\phi^{\rm (v)} +M_{14}^{(5)} \phi^{\rm (v)}+\dots,\\\label{Q2dot}
{\Z}_{,t} &=& M_{21}^{(2)} P_Z^{\rm (ii)} + M_{22}^{(4)} \Z^{\rm (iv)}
+ M_{23}^{(3)} P_\phi^{\rm (iii)} + M_{24}^{(3)} \phi^{\rm (iii)}+\dots,\\\label{P5dot}
{P_\phi}_{,t} &=& M_{31}^{(3)} P_Z^{\rm (iii)} + M_{32}^{(5)} \Z^{\rm(v)}
+ M_{33}^{(4)} P_\phi^{\rm (iv)} + M_{34}^{(4)} \phi^{\rm (iv)}+\dots,\\\label{Q5dot}
{\phi}_{,t} &=& M_{41}^{(3)} P_Z^{\rm (iii)} + M_{42}^{(5)} \Z^{\rm (v)}
+ M_{43}^{(4)} P_\phi^{\rm (iv)} + M_{44}^{(4)} \phi^{\rm (iv)}+\dots,
\end{eqnarray}
where the Roman numerals in the different superscripts represent radial derivatives
and the dots stand for terms with lower-order radial derivatives. In addition, the subindices of
the background-dependent $M$-coefficients denote positions in a matrix.
More precisely, $M_{ij}^{(k)}$ would correspond to the slot
$(i,j)$ of the matrix that multiplies the $k$th order radial derivative
of the column vector $(P_Z,\Z,P_\phi,\phi)^T$.
Before further analyzing this system of equations, let us particularize it to
the vacuum case and recover the results obtained in such a scenario by Moncrief \cite{Monc74}.

\subsection{The vacuum case: the Zerilli equation}

For a vacuum spacetime, all background and perturbative fluid variables
$\{\Pi_3,\Phi, Q_5, P_5\}$ disappear from our problem and
equations (\ref{P5dot}-\ref{Q5dot}) are empty. In this case,
the background gauge conditions we have chosen
at the beginning of this section, which implied $\Pi_1=0$,
also impose a vanishing value for the background momentum $\Pi_2$
due to the constraint (\ref{sphericalconstraint2}).
In addition, from background equations (\ref{pi1dot},\ref{pi2dot})
one gets the explicit form of the metric components,
\begin{equation}
\alpha=\frac{1}{a}=\sqrt{1-\frac{2 M}{r}}.
\end{equation}
In this way, one can solve equation (\ref{Q2dot}) to write down
the gauge-invariant momentum $P_Z$ in terms of the time derivative
of its conjugate variable $\Z$,
\begin{equation}\label{PZvacuumZt}
P_Z=\frac{2r^6\Lambda}{l(l+1)(2r\Lambda+6M)^2}\Z{}_{,t}.
\end{equation}
For convenience, we define the rescaled variable,
\begin{equation}
{{\Zer}} :=-\frac{r^3}{a}
\frac{\Z}{2r\Lambda+6 M},
\end{equation}
which, inverting the canonical transformations that have
been performed in the previous section, can be expressed in terms of
the initial harmonic coefficients as
\begin{equation}\label{rescaled}
\Zer = \frac{(r-2 M)}{3M+\Lambda r}\left\{
r{H}_{2} -r^2{K}_{,r}-l(l+1)h_1\right\}
+r{K}
+\frac{1}{2} l(l+1) r {G}.
\end{equation}

Finally, introducing relation (\ref{PZvacuumZt}) in equation (\ref{P2dot}),
the Zerilli equation is obtained,
\begin{equation}\label{zereqtortoise}
\left(1-\frac{2M}{r}\right)^{-1}\left(-\frac{\partial^2{\Zer}}{\partial
t^2} +\frac{\partial^2{\Zer}}{\partial r^*{}^2}\right)
-V_{\rm Z}\,\Zer=0,
\end{equation}
where we have made use of the tortoise coordinates $(t,r^*)$, with $r^*=r+2M
\ln(\frac{r}{2M}-1)$, and the potential is given by,
\begin{equation}\label{Zpotential}
V_{\rm Z}:=\frac{l(l+1)}{r^2}
-\frac{6M}{r^3}\frac{r^2\Lambda(\Lambda+2)+3M(r-M)}{(r\Lambda+3M)^2}.
\end{equation}
Therefore, the gauge-invariant combination $\Zer$ (\ref{rescaled})
reduces to the Zerilli variable when particularized to vacuum.

\subsection{Simplification of the evolution system}

The highest radial derivative that appears in the evolution equations (\ref{P2dot}--\ref{Q5dot})
is that of the $\Z$ variable. More precisely, in Eq. (\ref{P2dot}) it is of sixth order, whereas in Eqs. (\ref{P5dot}) and
(\ref{Q5dot}) it is of fifth order. Nonetheless in Eq. (\ref{Q2dot}) only radial derivatives of the
variable $\Z$ up to fourth order appear. Thus, this latter equation can be derived with respect to
$r$ in order to replace those higher-radial derivatives of $\Z$, which appear in other equations,
with lower-order derivatives of different variables and a time derivative of $\Z$.
In fact, it turns out that it is possible to perform a change of variables
so that the highest radial derivative in the equations is of third order only. Let us define
the new variables
\begin{eqnarray}\label{transformationxi1}
\xi_1&=&B_1{P_Z} + B_2{\Z}_{,rr} + B_3{\Z}_{,r},\\
\xi_2&=&\Z,\\
\xi_3&=& B_4{P_\phi} + B_5{\Z}_{,r},\\\label{transformationxi4}
\xi_4&=& B_6{\phi} + B_7{\Z}_{,r},
\end{eqnarray}
where the $B_i$ coefficients are given by,
\begin{eqnarray}
B_1&:=&  \frac{2\alpha^2}{\lambda^2} \left(\frac{r\Pi_3}{a}\right)^6
\frac{\Pi_2}{aD},\\
B_2&:=& \frac{2\alpha^2}{\lambda^2}
\left(\frac{r^4\Pi_2\Pi_3^2}{a^4D}\right)^2,\\
B_3&:=&-\frac{\alpha^2}{2\lambda^2}
\left(\frac{\Pi_3 r^3 }{a^4 D}\right)^2\Pi_2
\Bigg\{4r^3\Pi_3^3 \Phi_{,\rho\rho}+r\left[r^2\Pi_3^2+l (l+1)\right]
\Pi_2(\Phi_{,\rho})^2
+a^2r^3\Pi_2\Pi_3^4\nonumber\\
&+&\frac{4r^2}{D}\left(D\Pi_3^2\right)_{,\rho}\Pi_2
+r\left[3 a^2 l (l+1)-8\right]\Pi_2\Pi_3^2
+\frac{4}{r} l(l+1) \left[a^2\left(l^2+l+1\right)-3\right]\Pi_2\Bigg\},
\\
B_4&:=& \frac{2\alpha}{\lambda}\left(\frac{r^2\Pi_3}{a^2}\right)^2
\frac{\Pi_2}{D},\\
B_5&:=& \frac{\alpha r}{\lambda}
\left(\frac{r^2}{a^3 D}\right)^2 \Pi_2
\Big\{
\left[\left(l^2+l+2\right) a^2+2\right]\Phi_{,\rho}\Pi_2
+2 ra^2D\Pi_3+2 r \Phi_{,\rho\rho}\Pi_2
\Big\}\!,\\
B_6&:=&-B_4,\\
B_7&:=& -\frac{2\alpha r}{\lambda}\left(\frac{r }{a}\right)^3
\left(\frac{\Pi_2}{aD}\right)^2\Pi_3.
\end{eqnarray}
The system of equations (\ref{P2dot}--\ref{Q5dot}) is then rewritten in
the following way:
\begin{equation}
\frac{\partial \xi_i}{\partial t}=\sum_{j=1}^4\sum_{k=0}^3 A^{ij}_{k} \frac{\partial^k \xi_j}{\partial r^k},
\end{equation}
or, by defining the column vector $\vec{\xi}:=(\xi_1,\xi_2,\xi_3,\xi_4)^T$, in matrix notation,
\begin{equation}\label{system1}
\frac{\partial \vec{\xi}}{\partial t}= \sum_{k=0}^3 A_{k} \frac{\partial^k \vec{\xi}}{\partial r^k}.
\end{equation}
The first row of the matrix corresponding to third-order radial derivatives is zero, that is $A^{1j}_{3}=0$,
and thus only up to second-order derivatives of $\xi_1$ appears in the equations.
Note that the global differential order of the system of equations has been reduced
in three (from up to sixth-order radial derivatives to up to just third order)
by the above transformation (\ref{transformationxi1}-\ref{transformationxi4}).

In order to further simplify this set of equations,
one could perform another change of variables that takes
the matrix coefficients of the higher-order derivatives, in this case $A_{3}$,
to its Jordan form. In the axial case such a transformation converted a set
of two equations of second-order into a new set of an equation of second-order
and another with no radial derivatives \cite{BrMa09}.

In the present case,
all four eigenvalues of the matrix $A_{3}$ are zero and its corresponding
Jordan form is,
\begin{equation}
J=\left(
\begin{array}{cccc}
 0 & 1 & 0 & 0 \\
 0 & 0 & 0 & 0 \\
 0 & 0 & 0 & 1 \\
 0 & 0 & 0 & 0 \\
\end{array}
\right).
\end{equation}
As it is usual, a similarity transformation, implemented by a matrix $S$, relates the matrix $A_3$ with its Jordan form $J$,
\begin{equation}
J=S^{-1}A_3S.
\end{equation}
With this $S$ matrix at hand, one can define the normal coordinates $\vec{\omega}$ as
\begin{equation}
\vec{\omega}:= S^{-1}\vec{\xi}.
\end{equation}
And the equations of motion for these new variables will take the following form:
\begin{equation}\label{system2}
\frac{\partial\vec{\omega}}{\partial t}=J\,\frac{\partial^3\vec{\omega}}{\partial r^3}+ \sum_{k=0}^2\Omega_k \frac{\partial^k\vec{\omega}}{\partial r^k}.
\end{equation}
It is easy to see that these $\Omega$ matrices can be written in terms of $A_k$
matrices in combination with derivatives of the similarity matrix $S$ in the following way,
\begin{eqnarray}
\Omega_2&= &S^{-1} (3 A_3S_{,r}+A_2S),\\
\Omega_1&= &S^{-1} (3 A_3S_{,rr}+2A_2S_{,r}+A_1S),\\
\Omega_0&= &S^{-1} (A_3S_{,rrr}+A_2S_{,rr}+A_1S_{,r}+A_0S-S_{,t}).
\end{eqnarray}
Unfortunately, these matrices are quite involve. In particular, and as opposed to what happened in
the axial case, all the components of the matrix $\Omega_2$ are non-vanishing and thus
second radial derivatives of all variables appear in all equations of the system (\ref{system2}).
Therefore the only advantage of the normal variables $\vec{\omega}$ with respect
to the initial variables $\vec{\xi}$ lies in the simplicity of the third-order radial derivative
terms, which only appear in two of the variables. Nonetheless, this advantage might not
be so relevant if one considers that our final interest is to obtain the perturbed metric
in terms of the variables for which we solve. Reconstructing the perturbed metric
from the $\vec{\omega}$ variables implies another change of variables more than
reconstructing it from $\vec{\xi}$ variables. In addition, $\Omega_k$ matrices are more
involve than their $A_k$ counterparts for all $k\leq 2$.
The lengthy expressions of $A_k$ and $\Omega_k$ prevents us from providing
them here explicitly. Even so, these matrices are available from the author by request.

\section{Conclusions}\label{sec:conclusions}

In this paper a generalization of the Zerilli master variable for a specific spherical but
dynamical background spacetime has been presented. In order to factorize and remove the
angular dependence from the equations of motion, a decomposition on tensor spherical
harmonics of the polar part of different perturbative variables has been performed.
At linearised level each harmonic coefficient, characterized by $(l,m)$
angular numbers, decouple from the rest due to the symmetry of the background.
The perturbative problem has then been formulated on a Hamiltonian framework, which shows very clearly
the dynamical role of each object. As it is well known, the second variation of the
Einstein-Hilbert action provides an action functional for the perturbative variables. In this way, an effective
Hamiltonian can be defined for linearised variables, which is given as a linear combination of a physical
Hamiltonian and four constraints, which are the linearised constraints of general relativity.
These constraints are the generators of gauge transformations. Thus, in order to construct gauge-invariant
master variables, one can perform a canonical transformation so that four of the new variables are
equal to the constraints. In this way, for the considered background spacetime, one obtains two pairs of variables
that encode the complete physical information of the problem: one pair corresponds to the polar
mode of the gravitational wave and the other one to the scalar matter degree of freedom.

In order to be admissible,
three conditions have been requested to this canonical transformation, which has been performed
as five subsequent transformations to provide a clear view of each step.
First, they should be algebraic transformations so that they do not involve any integration
and can be performed explicitly. Second, they should not require dividing by any
background object that could vanish. And third, they should have a well defined vacuum limit
so that we obtain a generalization of the Zerilli variable.
This latter in particular implies that the perturbative matter degrees of freedom
can not be used to solve for the perturbative constraints, since such a transformation would
not be well defined in the limit that the perturbations of the scalar field vanish.
The full transformation that has been proposed here completely fulfills the first
and third criteria, but it is unclear whether it also satisfies the second one.
In the last (fifth) transformation it turned out to be necessary to divide by
the background coefficient $D$ defined in (\ref{background_coefficient}). This background coefficient
is positive definite in the Schwarzschild case (\ref{deflambda}), but it is difficult to assert
something about it in the general dynamical case.

In principle the master variable that has been found is not unique, since
there is apparently much freedom in the canonical transformations that one could perform.
Nonetheless, the three imposed criteria reduce considerably this freedom. In particular note that,
once the canonical transformation (\ref{trans1first}--\ref{trans1last}) is performed, these criteria
almost completely single out the subsequent transformations.
More specifically, the momenta $\pi_3$ and $\pi_4$ are the only variables that 
can be used to solve for the constraints (\ref{deltaHpolar}) and (\ref{deltaHsubrho}) respectively.
This leads to the form (\ref{intermediatedeltaH}) of the linearised Hamiltonian constraint.
In that expression none of the variables can be used to solve the constraint
algebraically. Thus, next parametrized transformation (\ref{trans4first}--\ref{trans4last})
is performed in order to concentrate all derivatives in a unique full derivative.
The term inside this full derivative is then promoted to one of the basic variables.
Finally, in expression (\ref{Ham4}) the variable $\check k_1$, which appears with no
derivatives, is used to solve for the Hamiltonian constraint. Note that $\check k_5$
could also be used to solve for that constraint, but then this transformation
would not have a well defined limit when the perturbative matter degrees of freedom vanish
and would then violate the third criterion.

Finally, the evolution equations obeyed by the master variables have been obtained.
The differential order, in radial derivatives, was initially seven. Nonetheless, it is
possible to redefine new master variables so that the highest radial derivative is
of third order. Hence contrary to the Zerilli variable, which fulfills a wave equation, these
master variables obey equations of higher (radial) order. This could be surprising
since the equation of motion of a linear combination of objects, which
obey second-order differential equations, is obviously of second-order in terms of the objects themselves.
Nonetheless, the fact that these objects are expressed in terms of the combination itself (and its radial derivatives)
can increase the differential order of the equation. This is exactly what happens in this case.
Unfortunately the obtained equations are quite involved and it is not clear if they could be of practical use.
Nevertheless, the fact that the mentioned canonical transformations can be performed algebraically and completely
decouple the gauge from the physical degrees of freedom, turns out to be a relevant
result in itself, which could pave the way for the construction of a polar master variable
for any dynamical spherically symmetric background.

\acknowledgments

The author would like to thank Jos\'e M. Mart\'in-Garc\'ia for proposing this problem,
many interesting discussions and continuous encouragement; and Guillermo A. Mena Marug\'an
for discussions in the initial stages of the project.
This work is supported by Projects IT592-13 of the Basque Government and FIS2012-34379
of the Spanish Ministry of Economy and Competitiveness.




\begin{thebibliography}{99}

\bibitem{Monc74} V. Moncrief, Annals of Phys. {\bf 88}, 323 (1974).
\bibitem{ReWh57} T. Regge and J. A. Wheeler, Phys. Rev. {\bf 108}, 1063 (1957).
\bibitem{Zeri70} F. J. Zerilli, Phys. Rev. Lett. {\bf 24}, 737 (1970).
\bibitem{Monc74b} V. Moncrief, Annals of Phys. {\bf 88}, 343 (1974).
\bibitem{Monc75} V. Moncrief, Phys. Rev. D {\bf 9}, 2707 (1974); {\bf 10}, 1057 (1974); {\bf 12}, 1526 (1975).
\bibitem{CPM78} C. T. Cunningham, R. H. Price, and V. Moncrief, Astrophys. J. {\bf 224}, 643 (1978).
\bibitem{Gund93} C. Gundlach, Class. Quant. Grav. {\bf 10}, 1103 (1993).
\bibitem{Lan94} D. Langlois, Class. Quant. Grav. {\bf 11}, 389 (1994).
\bibitem{PBMT10} E. Pazos, D. Brizuela, J. M. Mart\'in-Garc\'ia, and M. Tiglio, Phys. Rev. D {\bf 82}, 104028 (2010).
\bibitem{GeSe79} U. H. Gerlach and U. K. Sengupta, Phys. Rev. D {\bf 19}, 2268 (1979); {\bf 22}, 1300 (1980).
\bibitem{GuMa00} C. Gundlach and J. M. Mart{\'\i}n-Garc{\'\i}a, Phys. Rev. D {\bf 61}, 084024 (2000).
\bibitem{MaPo05} K. Martel and E. Poisson, Phys. Rev. D {\bf 71}, 104003 (2005).
\bibitem{NaRe05} A. Nagar and L. Rezzolla, Class. Quant. Grav. {\bf 22}, R167 (2005); Erratum-ibid {\bf 23}, 4297 (2006).
\bibitem{SaTi00} O. Sarbach and M. Tiglio, Phys. Rev. D {\bf 64}, 084016 (2001).
\bibitem{MoSa03} C. Moreno and O. Sarbach, Phys. Rev. D {\bf 67}, 024028 (2003).
\bibitem{Sei90} E. Seidel, Phys. Rev. D {\bf 42}, 1884 (1990).
\bibitem{MaGu01} J. M. Mart\'\i n-Garc\'\i a and C. Gundlach, Phys. Rev. D {\bf 64}, 024012 (2001).
\bibitem{MMR07} M. Mars, F. C. Mena, and R. Vera, Class. Quant. Grav. {\bf 24}, 3673 (2007).
\bibitem{Mar05} M. Mars, Class. Quant. Grav. {\bf 22}, 3325 (2005).
\bibitem{Muk00} S. Mukohyama, Class. Quant. Grav. {\bf 17}, 4777 (2000).
\bibitem{BMSK10} D. Brizuela, J. M. Mart\'in-Garc\'ia, U. Sperhake, and K. D. Kokkotas, Phys. Rev. D {\bf 82}, 104039 (2010).
\bibitem{ReVe14} B. Reina and R. Vera, arXiv:1412.7083 [gr-qc].
\bibitem{BrMa09} D. Brizuela and J. M. Mart\'in-Garc\'ia, Class. Quant. Grav. {\bf 26}, 015003 (2009).
\bibitem{Taub} A. Taub, Commun. Math. Phys. {\bf 15}, 235 (1969).
\bibitem{BMM06} D. Brizuela, J. M. Mart\'\i n-Garc\'\i a and G. A. Mena Marug\'an,
Phys. Rev. D {\bf 74}, 044039 (2006).


\end{thebibliography}
\end{document}